\documentclass[aps,prl,floatfix,twocolumn,superscriptaddress]{revtex4}
\usepackage{float}
\usepackage{makeidx}
\usepackage{graphicx,amsmath}
\usepackage{colordvi}
\usepackage[dvipsnames]{xcolor}
\usepackage{ulem}
\allowdisplaybreaks

\emergencystretch=\maxdimen
\usepackage{mathrsfs}

\begin{document}

\title{$\pi$-flux Dirac bosons and topological edge excitations \\
in a bosonic chiral $p$-wave superfluid}

\author{Zhi-Fang Xu}
\email{zfxu83@gmail.com}
\affiliation{MOE Key Laboratory of Fundamental Physical Quantities Measurements, School of Physics, Huazhong University of Science and Technology, Wuhan 430074, China}
\author{Li You}
\affiliation{State Key Laboratory of Low Dimensional Quantum Physics, Department of Physics, Tsinghua University, Beijing 100084, China}
\affiliation{Collaborative Innovation Center of Quantum Matter, Beijing 100084, China}
\author{Andreas Hemmerich}
\affiliation{Institut f\"ur Laser-Physik, Universit\"at Hamburg, Luruper Chaussee 149, 22761 Hamburg, Germany}
\affiliation{Wilczek Quantum Center, Zhejiang University of Technology, Hangzhou 310023, China}
\author{W. Vincent Liu}
\email{wvliu@pitt.edu}
\affiliation{Department of Physics and Astronomy, University of Pittsburgh, Pittsburgh, Pennsylvania 15260, USA}
\affiliation{Wilczek Quantum Center, Zhejiang University of Technology, Hangzhou 310023, China}
\affiliation{Center for Cold Atom Physics, Chinese Academy of Sciences, Wuhan 430071, China}

\begin{abstract}
We study the topological properties of elementary excitations in a staggered $p_x \pm i p_y$ Bose-Einstein condensate realized in recent orbital optical lattice experiments. The condensate wave function may be viewed as a configuration space variant of the famous $p_x+ ip_y$ momentum space order parameter of strontium ruthenate superconductors. We show that its elementary excitation spectrum possesses Dirac bosons with $\pi$ Berry flux. Remarkably, if we induce a population imbalance between the $p_x+ip_y$ and $p_x-ip_y$ condensate components, a gap opens up in the excitation spectrum resulting in a nonzero Chern invariant and topologically protected edge excitation modes. We give a detailed description on how our proposal can be implemented with standard experimental technology.
\end{abstract}

\pacs{67.85.-d, 03.75.Kk, 03.75.Lm, 03.65.Vf}

\maketitle

{\it Introduction.} In quantum gases, the current focus on topological features so far has been directed towards fermionic atoms, whose phase diagrams often are easily obtained, thanks to extensive knowledge derived from analogous electron models~\cite{Hasan2010,Qi2011}. Consorted efforts in cold atom physics aim at simulating famous electronic topological models that are non-interacting but possess geometric phases~\cite{Aidelsburger2013, Miyake2013, Aidelsburger2015, Jotzu2014, Mancini2015, Stuhl2015} in recent years. Synthetic magnetic flux or spin-orbit coupling are achieved by experimental techniques of laser-induced tunneling~\cite{Lin2011, Aidelsburger2013, Miyake2013, Aidelsburger2015, Mancini2015, Stuhl2015} and lattice modulation techniques~\cite{Struck2011, Parker2013, Jotzu2014}. With regard to interacting systems, fermionic superfluid (superconducting) phases with salient orbital symmetry (such as $p$ or $d$ wave)~\cite{Volovik2003, Kallin2012} have been actively searched for decades. They support edge-state fermions featuring non-Abelian braiding statistics~\cite{Nayak2008} within the energy gap opened up along the Fermi surface.

Recently, the search for similar phenomena in bosonic systems, which are easier to implement with cold atomic gases, attracts broad interest. Several theoretical ideas ~\cite{Engelhardt2015, Furukawa2015, Bardyn2015} have been put forward  to realize  topological bosonic elementary excitations. One can utilize bosons condensed in the minimum of the lowest single-particle Bloch band, which is endowed with topological character by means of a synthetic gauge field~\cite{Engelhardt2015, Furukawa2015}. An alternative approach is based on the formation of a background vortex-lattice condensate with optical phase imprinting~\cite{Bardyn2015}. Both types of proposals imply significant experimental complexity.

In this Letter, we find that because of the interplay of interaction and orbital symmetry, topological elementary excitations naturally  occur in a staggered chiral bosonic $p_x\pm ip_y$ superfluid discovered in a series of recent experiments at Hamburg~\cite{Wirth2011, Olschlager2013, Kock2015}. We show that it supports Dirac bosons with $\pi$ Berry flux in a higher lying branch of excitation spectrum. Surprisingly, we find that via adjusting the population imbalance of the $p_x+ip_y$ and $p_x-ip_y$ components, a topological phase transition occurs, accompanied by a bulk gap opening up near the Dirac cones. This leads to finite energy in-gap edge excitations for the finite system.  Most strikingly, the topologically protected edge excitations are generated by the background chiral superfluid, which distinguishes this work from others that rely on engineering of the single-particle band structure. Below, we will discuss in detail how the required population imbalance tuning can be readily achieved with standard experimental techniques.

We thus realize a bosonic counterpart of topological chiral fermionic superfluidity, which is reminiscent of the famous example of topological superconductivity: the fermionic $p_x + ip_y$ state, believed to occur in strontium ruthenates~\cite{Maeno1994,Mackenzie2003}. Their shared features include (1) both bosonic and fermionic superfluid states have nonzero global orbital angular momentum, and (2) both support topologically protected in-gap edge excitations. However, a fundamental difference exists. The fermionic state has a bulk gap above the ground state and supports zero modes at material edges or in vortex cores, while the bosonic superfluid has gapless Nambu-Goldstone modes immediately above the ground state due to the continuous global U(1) gauge symmetry breaking and the topological edge states occur inside a bulk gap at high energies.

\begin{figure*}
\centering
\includegraphics[width=0.75\linewidth]{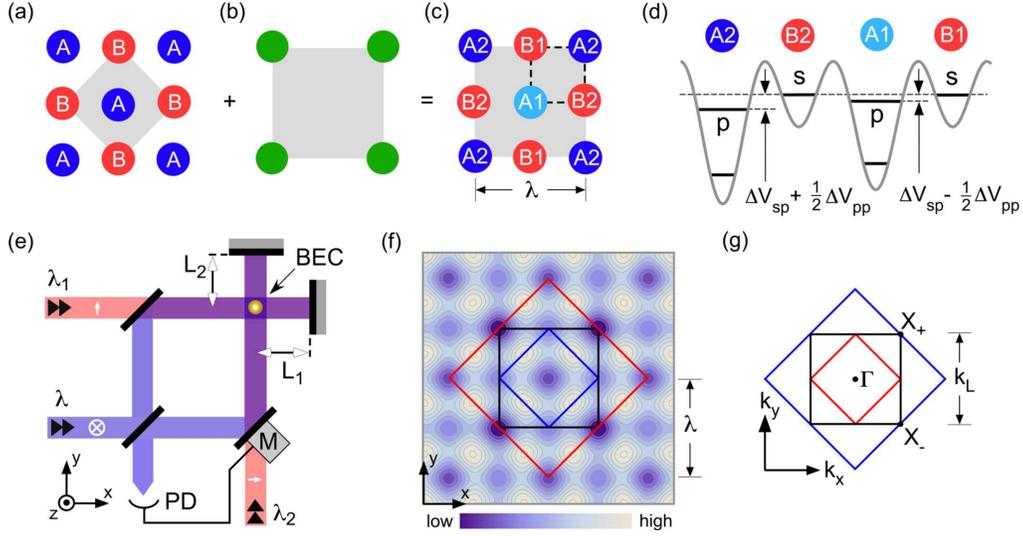}
\caption{ (color online). (a) The checkerboard potential $V_H$ comprises two classes of wells denoted $A$ and $B$. (b) Geometry of the added square lattice $V_S$ for the simplest case $\eta=0$. (c) The complete lattice $V=V_H + V_S$ comprises four classes of sites denoted as $A1$, $A2$, $B1$, and $B2$. The gray rectangles in (a), (b), (c) indicates the unit cells of $V_H$, $V_S$, and $V$, respectively. (d), The potential $V$ is shown along the dashed square in (c). The $s$ and $p$ orbitals accounted for in the tight binding model are indicated. (e) Laser beams at wavelengths $\lambda$,  $\lambda_1$ and $\lambda_2$ are coupled to the interferometric setup for the formation of $V$. Their linear polarizations (orthogonal or parallel with respect to the $xy$ plane) are indicated by white arrows. The phase $\beta$ in Eq.~(\ref{checkerboardOL}) is servo-controlled by a movable mirror (M) and a photo detector (PD), in order to tune $\Delta V_{\rm sp}$ in Eq.~(\ref{hamiltonian}). (f), Contour plot of the lattice potential $V=V_H + V_S$. The inner blue and middle black solid squares denote the unit cells of the lattice potential $V_H$ and $V$, respectively. The outer red solid square denotes the unit cell of the predicted order parameter. (g), The first Brillouin zones corresponding to the unit cells shown in (f).}
\label{fig1}
\end{figure*}

{\it Model.} We employ a generalized version of the lattice geometry employed in earlier experiments~\cite{Wirth2011,Olschlager2013,Kock2015}  given by the potential $V(x,y)=V_{H}(x,y)+V_S(x,y)$ with
\begin{eqnarray}
V_H(x,y)&=& -V_1\left|\cos(k_L x)+e^{i\beta}\cos(k_L y)\right|^2, \\ \nonumber
V_S(x,y)&=&V_2\left[\cos^2(\frac{1+2 \eta}{2} k_L x)+\cos^2(\frac{1+2 \eta}{2} k_L y)\right].
\label{checkerboardOL}
\end{eqnarray}
Here, $V_H$ is the original lattice of the Hamburg experiments, which provides deep and shallow wells, denoted by $A$ and $B$ (see Fig.~\ref{fig1}(a)). This lattice is formed by two interfering optical standing waves aligned along the $x$- and $y$-axes, respectively, with linear polarizations parallel to the $z$-axis, wavelength $\lambda$  (with $k_L=2\pi/\lambda$) and relative phase $\beta$, as detailed in Refs.~\cite{Wirth2011,Olschlager2013,Kock2015,Hemmerich1991}. We superimpose an additional weak ($|V_2| \ll |V_1|$) conventional square lattice potential $V_S$ as used in many experiments, with wave number $\frac{1+2 \eta}{2} k_L$ and $\eta \in \{0,1,2,...\}$ (see Fig.~\ref{fig1}(b) for the case $\eta=0$). $V_S$ acts to subdivide the deep $A$-sites into two classes ($A1,A2$) of slightly different depth.

In the tight-binding regime, by taking into account $p_x$ and $p_y$ orbitals in the deep ($A1,A2$) and $s$ orbitals in the shallow wells ($B1,B2$), we model our system by the extended Hubbard Hamiltonian $\hat{H}=\hat{H}_0+\hat{H}_{\rm int}$, where
\begin{eqnarray}
    \hat{H}_0&=&-J_{\rm sp}\sum\limits_{\mathbf{r},\mu}\left[
    \hat{b}_{\mu,\mathbf{r}}^{\dagger}\hat{b}_{s,\mathbf{r}+\mathbf{e}_{\mu}} -\hat{b}_{\mu,\mathbf{r}}^{\dagger}\hat{b}_{s,\mathbf{r}-\mathbf{e}_{\mu}} +\rm h.c.\right] \nonumber\\
    &&+J_{\parallel}\sum\limits_{\mathbf{r},\mu}\left[
    \hat{b}_{\mu,\mathbf{r}}^{\dagger}\hat{b}_{\mu,\mathbf{r}+\mathbf{e}_x+\mathbf{e}_y} +\hat{b}_{\mu,\mathbf{r}}^{\dagger}\hat{b}_{\mu,\mathbf{r}+\mathbf{e}_x-\mathbf{e}_y} +\rm h.c. \right] \nonumber\\
    &&+J_{\perp}\sum\limits_{\mathbf{r},\mu}\left[
    \hat{b}_{\mu,\mathbf{r}}^{\dagger}\hat{b}_{\bar{\mu},\mathbf{r}+\mathbf{e}_x+\mathbf{e}_y} -\hat{b}_{\mu,\mathbf{r}}^{\dagger}\hat{b}_{\bar{\mu},\mathbf{r}+\mathbf{e}_x-\mathbf{e}_y} +\rm h.c. \right] \nonumber\\
    &&+\sum\limits_{\mathbf{r}} \left[-\Delta V_{\rm sp}+\frac{1}{2}\Delta V_{\rm pp}(-)^{r_x+r_y}\right]\hat{n}_{p,\mathbf{r}},
    \label{hamiltonianSP}\\
    \hat{H}_{\rm int}&=&\frac{U_s}{2}\sum\limits_{\mathbf{r}}\hat{n}_{s,\mathbf{r}+\mathbf{e}_x} (\hat{n}_{s,\mathbf{r}+\mathbf{e}_x}-1)\nonumber\\
    &&+\frac{U_p}{2}\sum\limits_{\mathbf{r}}\left\{
    \hat{n}_{p,\mathbf{r}}\left[\hat{n}_{p,\mathbf{r}}-\frac{2}{3}\right]
    -\frac{1}{3}\hat{L}_{z,\mathbf{r}}^2\right\}.
\label{hamiltonian}
\end{eqnarray}
Here, $\hat{b}_{s,\mathbf{r}'}$ and $\hat{b}_{\mu,\mathbf{r}'}$are bosonic annihilation operators for $s$ and $p_{\mu}$ ($\mu=\{x,y\}$) orbitals at site $\mathbf{r}'$. The vector $\mathbf{r}=(r_x,r_y)\lambda$ with integer $r_x$ and $r_y$ denotes the position of $p$ orbitals, while $s$ orbitals are located at $\mathbf{r}+\mathbf{e}_{\mu}$ with $\mathbf{e}_{x}=(\lambda/2,0)$ and $\mathbf{e}_{y}=(0,\lambda/2)$. $\bar{\mu}=\{\bar{x},\bar{y}\}=\{y,x\}$. The number and the on-site angular momentum operators are respectively given by $\hat{n}_s=\hat{b}_s^{\dagger}\hat{b}_s$, $\hat{n}_p=\hat{b}_x^{\dagger}\hat{b}_x+\hat{b}_y^{\dagger}\hat{b}_y$, and $\hat{L}_z=-i(\hat{b}_x^{\dagger}\hat{b}_y-\hat{b}_y^{\dagger}\hat{b}_x)$. The $\Delta V_{\rm pp}$ term arises from the added potential $V_S$, which induces a staggered on-site energy shift for $p$ orbitals at $A$ sites.

\begin{figure}[tbp]
\centering
\includegraphics[width=\linewidth]{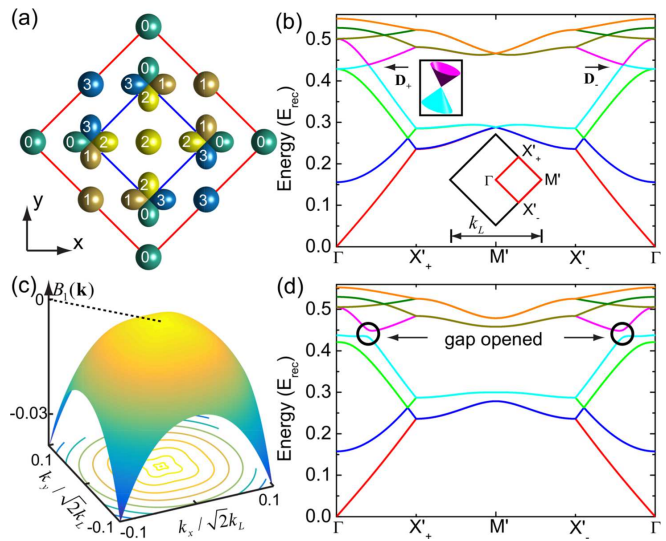}
\caption{ (color online). (a) The schematic picture of the  order parameter in one unit cell denoted by the outer solid square. The integer number $\nu$ denotes the phase of the orbital wavefunction as $\nu\,\pi/2$. Specifically, $p$ orbitals are located at the four corners of the inner square, while other sites are occupied by the $s$ orbitals. (b) Excitation spectra for $V_S=0$ along the high-symmetric lines in the first Brillouin zone marked by the red diamond in the lower inset. Two Dirac points are denoted by $D_+$ and $D_-$. The parameters used in numerics are $J_{\rm sp}=0.12\, E_{\rm rec}$, $J_{\parallel}=0.07\, J_{\rm sp}$, $J_{\perp}=0.15\, J_{\rm sp}$, $\Delta V_{\rm sp}=0.3\, J_{\rm sp}$, $U_s=0.24\, E_{\rm rec}$, $U_p=0.12\, E_{\rm rec}$, and $\Delta_{\rm pp}=0$. The mean boson density corresponds to one particle per orbital. Here, only the lowest $8$ bands are shown. (c) Berry curvature of the first band of elementary excitation. (d) The same as (b) except a bias square optical lattice potential $V_S$ is turned on to induce a population imbalance between $p_x+ip_y$ and $p_x-ip_y$ components, which opens a gap around the Dirac points leading to topological elementary excitations and nonzero Berry curvature for the phonon mode shown in (c). Here, the only difference in parameters used in numerics is $\Delta_{\rm pp}=3\, \Delta_{\rm sp}$. }
\label{fig2}
\end{figure}

For the single-particle Hamiltonian $H_0$, the band minima are located at $X_{\pm}=(1,\pm1) k_L/2$ independent of $V_S$. This yields a finite-momentum condensate for weakly interacting bosons. The corresponding unit cell of the order parameter is enlarged by a factor of $\sqrt{2}$ and rotated by $45^{\circ}$ with respect to that of the lattice potential $V=V_H + V_S$, hence is composed of four $A$ and four $B$ sites, as illustrated by the red solid square in Fig.~\ref{fig1}(f). Ferro-orbital interaction favors a special phase correlation between the two band minima, leading to a staggered on-site orbital angular momentum distribution with the relative phase between $p_x$ and $p_y$ orbitals fixed at $\pm \pi/2$~\cite{Isacsson2005,Liu2006}. Figure~\ref{fig2}(a) shows the schematic representation of the ground-state order parameter. While previous studies concentrated on the chiral nature of such a $p$ orbital bosonic superfluid, the present study investigates the elementary excitations on top of it instead. Remarkably, we find that interactions among bosons not only generate a ground state with chiral orbital order as is understood and experimentally confirmed, but also give rise to topological quasiparticle excitations to be described in the following.

{\it Topological elementary excitations.} We follow the standard number-conserving approach to obtain the excitation spectra~\cite{suppl}. Firstly, we focus on the symmetry of the Bogoliubov-de Gennes (BdG) Hamiltonian $\mathcal{H}(\mathbf{k})$. The presence of the staggered chiral orbital order breaks both time reversal and original lattice translation symmetries. Nevertheless, for $V_S=0$, via choosing a specific global phase of the order parameter $\mathcal{H}(\mathbf{k})$ is invariant under the combined symmetry operations of time reversal $\mathcal{T}$ and lattice translations $\Theta_{\pm}$ along the $x\pm y$ directions by $\lambda/\sqrt{2}$: $\Theta_\pm \mathcal{T} \mathcal{H}(\mathbf{k})(\Theta_\pm\mathcal{T}) ^{-1} = \mathcal{H}(-\mathbf{k})$. The combined
symmetries are described by anti-unitary operators $\Theta_\pm \mathcal{T}$ and are broken for $V_S\ne0$. As will become clear later, this property plays an important role in the topological classification. The Hamiltonian is then diagonalized by a paraunitary matrix $T_{\mathbf{k}}$ as $T^{\dagger}_{\mathbf{k}}\mathcal{H}(\mathbf{k})T_{\mathbf{k}}=E_{\mathbf{k}}$, with eigenvalues contained in the diagonal terms of the diagonal matrix $E_{\mathbf{k}}$. To characterize the topological feature of the bosonic excitations, we follow the prescriptions of Refs.~\cite{Avron1983,Shindou2013} to define a topological invariant for the $j$-th excitation band in the reduced first Brillouin Zone (RFBZ) denoted by the red solid line in Fig.~\ref{fig1}(g) as
\begin{eqnarray}
 \mathcal{C}_j=\frac{1}{2\pi}\int_{\rm RFBZ}d^2\mathbf{k}B_j(\mathbf{k}),
 \label{topologicalinvariant}
\end{eqnarray}
where the Berry curvature $B_j$ and Berry connection $A_{j,\mu}$ are defined as $B_j(\mathbf{k})\equiv\partial_{k_x}A_{j,y}(\mathbf{k})-\partial_{k_y}A_{j,x}(\mathbf{k})$ and $A_{j,\mu}(\mathbf{k})\equiv i\text{Tr}[\Gamma_jT^{-1}_{\mathbf{k}}\tau_z(\partial_{k_{\mu}}T_{\mathbf{k}})]$. $\Gamma_j$ is a diagonal matrix with the $j$-th diagonal term equals to $1$ and other terms are 0, and $[\tau_z]_{ij}$ equals to $\delta_{ij}$ if $i=1,...,12$ and $-\delta_{ij}$ otherwise~\cite{suppl}.

We first consider the case of $V_S=0$ corresponding to the recently reported experiments~\cite{Wirth2011,Olschlager2013,Kock2015}. In this case,  $A1$ and $A2$ sublattice sites are equivalent, and therefore are equally populated in the $p$ orbital bands for the ground state. Fig.~\ref{fig2}(b) summarizes numerical results for the parameters at $J_{\rm sp}=0.12\, E_{\rm rec}$, $J_{\parallel}=0.07\, J_{\rm sp}$, $J_{\perp}=0.15\, J_{\rm sp}$, $\Delta V_{\rm sp}=0.3\, J_{\rm sp}$, $U_s=0.24\, E_{\rm rec}$, and $U_p=0.12\, E_{\rm rec}$, all in units of $E_{\rm rec}=\hbar^2k_L^2/(2m)$. Only the lowest $8$ energy bands along the high-symmetry lines are shown. These fully connected bands are gapped from the other $4$ topologically trivial higher energy bands due to the relatively larger contact interaction energy for the $s$ orbitals as compared to the $p$ orbitals.
Meanwhile, similar to a conventional Bose-Einstein condensate, the spontaneous breaking of U(1) phase symmetry guarantees the emergence of a phonon mode with linear dispersion close to $\mathbf{k}=\mathbf{0}$. We find the Berry curvature for this phonon mode to be negligible.

Surprisingly, an interesting feature in the excitations appears. Focusing on the $8$ bands shown in Fig.~\ref{fig2}(b), the 4-th and the 5-th bands are connected only at the four Dirac points located on the straight lines $\Gamma X'_{\pm}$ in momentum space. Two of them are denoted as $D_{\pm}$ and the other two correspond to their time-reversal points. Each of the Dirac points is associated with a $\pi$ Berry flux such that in its vicinity the excitation spectrum depends linearly on momentum. This feature, which is analogous to fermionic topological semimetals studied in Refs.~\cite{Neto2009,Sun2012}, may be measured by applying Aharonov-Bohm interferometry~\cite{Atala2013, Duca2015}.

\begin{figure}[tbp]
\centering
\includegraphics[width=\linewidth]{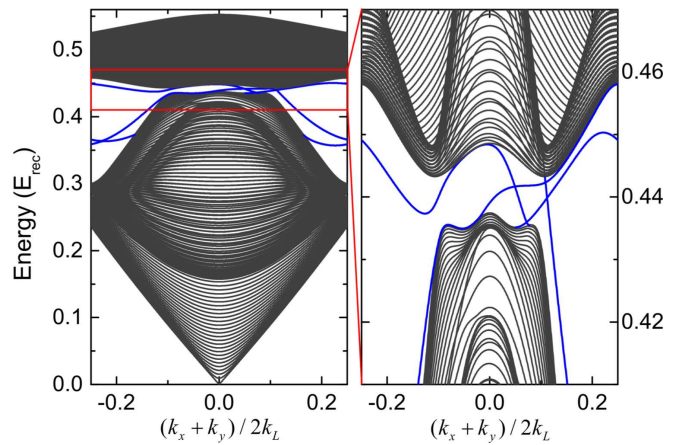}
\caption{ (color online). Elementary excitation spectra for a finite system with a cylinder geometry. A periodic (open) boundary condition is assumed in the $x+y$ ($x-y$) direction. One unit cell of the finite system contains 40 copies of the unit cell of the order parameter. The lattice potential and interaction parameters are the same as that used in Fig.~\ref{fig2}(d). Solid black and blue lines denote the bulk excitation spectra and the chiral edge states, respectively.}
\label{fig3}
\end{figure}

Next, we consider the general case with $V_S\ne0$. The order parameter shows the same phase correlation among $s$ and $p$ orbitals as for $V_S=0$, which is illustrated in Fig.~\ref{fig2}(a). The nonzero bias potential $V_S$ causes the $A1$ and $A2$ lattice sites to be inequivalent and hence induces a population imbalance between them. No matter how small $V_S$ is, a topological phase transition occurs. This is because the combined time reversal and discrete lattice translation symmetry $\Theta_{\pm}\mathcal{T}$ is broken for both the Hamiltonian and the order parameter, leading to a topological bulk gap on all Dirac points, as shown in Fig.~\ref{fig2}(d). For $V_S\ne 0$, the total 12 excitation bands are separated into three groups.  The four {\it connected} middle bands (5th-8th) are isolated by gap from four lower bands shown in Fig.~\ref{fig2}(d) and four highest bands which are not shown. The summation of their topological invariants are well defined. Using Eq.~(\ref{topologicalinvariant}), we confirm that the middle four bands are topological with $|\sum_{j=5}^8\mathcal{C}_j|=2$. To further elucidate their topological behavior, we investigate the chiral edge states of the system for a cylinder geometry with a periodic (or open) boundary condition in the $x+y$ (or $x-y$) direction. For this finite system, the numerically obtained excitation spectra are plotted in Fig.~\ref{fig3}. It clearly shows that topological edge states appear within the energy gap, connecting the upper and lower bulk states. The existence of two chiral edge states on each edge is consistent with the topological invariant being equal to $2$~\cite{Hasan2010}.

Refocusing on the phonon mode, we notice a striking feature appearing when $V_S\ne 0$. Different from the case of $V_S=0$, the bias potential $V_S$ induces a population imbalance for $p$ orbitals on $A1$ and $A2$ sites, leading to a nonzero total orbital angular momentum together with a nonzero Berry curvature for the phonon mode. Specifically, as shown in Fig.~\ref{fig2}(c), $B_1(\mathbf{k})$ is zero at the $\Gamma$ point and negative everywhere else in the $\mathbf{k}$-space area displayed, a situation implicating the existence of a phonon Hall effect~\cite{Xiao2010}.

Based on the above results, we understand that the population imbalance between the clockwise and the counter-clockwise orbiting condensate components is key to the appearance of topological elementary excitations in the $p$ band bosonic superfluid. The underlining symmetry for characterizing this imbalance is $\Theta_{\pm}\mathcal{T}$. To verify that this is indeed generally true rather than being specific to the form of $V_S$ used, we apply two other methods capable of opening a bulk gap close to the Dirac points~\cite{suppl}. The first one is to introduce an asymmetry between the $s$-$p$ orbital hopping amplitudes along two orthogonal directions, e.g., along the $x$- and $y$-axis, which corresponds to adding the following term
$\hat{H}_{1}=-\Delta J_{\rm sp}\sum_{\mathbf{r}}\big[ \hat{b}_s^{\dagger}(\mathbf{r})\hat{b}_y(\mathbf{r}+\mathbf{e}_y) + \hat{b}_s^{\dagger}(\mathbf{r})\hat{b}_y(\mathbf{r}-\mathbf{e}_y)+\rm h.c.\big]$ to the system Hamiltonian.
The second one is to generate an on-site rotation described by the Hamiltonian $\hat{H}_2=i\Omega_z\sum_{\mathbf{r}}\big[\hat{b}_x^{\dagger}(\mathbf{r}) \hat{b}_y(\mathbf{r})-\hat{b}_y^{\dagger}(\mathbf{r}) \hat{b}_x(\mathbf{r})\big]$. Our numerical calculations indeed confirm that the first method creates a trivial gap because it does not break the  $\Theta_{\pm}\mathcal{T}$ symmetry, while the second method relying on the on-site rotation breaks $\mathcal{T}$, thus breaks $\Theta_{\pm}\mathcal{T}$, giving rise to similar topological excitation bands as shown in Figs.~\ref{fig2}(d) and \ref{fig3}.

{\it Experimental realization and detection.} A schematic experimental setup for investigating our proposed study is shown in Fig.~\ref{fig1}(e). The weak $V_S$ can be realized by including two additional laser beams with linear polarizations within the $xy$ plane and wavelengths $\lambda_n = \lambda' + \delta\lambda_n$ with $\lambda' \equiv 2\lambda/(1+2 \eta)$. Take $\lambda=1064\,$nm and $\eta=1$, $\lambda' \approx 709\,$nm, a wave length for which laser light is readily available. The small quantities $\delta\lambda_n, n\in \{1,2\}$ are chosen according to the distances $L_n$ in Fig.~\ref{fig1}(e), in order to precisely adjust the relative positions of both lattices. Assuming $L_1 = L_2 = 20\,$cm a shift of $V_S$ by half a lattice constant amounts to $c / \delta\lambda_n = 375\,$MHz with $c$ denoting the speed of light. Technically, it is easily possible to fix $c / \delta\lambda_n$ with a precision of a few MHz such that the relative positions of the lattices can be adjusted to better than a few nm~\cite{suppl}.
As discussed in Ref.~\cite{Li2014}, the bias potential may also induce a staggered on-site coupling between $p_x$ and $p_y$ orbitals. However, such a coupling is proportional to the second derivative of the bias lattice potential, which is negligible for the weak lattice potential required here, e.g., $\Delta V_{\rm pp}\sim J_{\rm sp}$. The topological features of the elementary excitations can be experimentally measured via coherently transferring a small portion of the condensate into an edge mode by stimulated Raman transitions~\cite{Ernst2010,suppl}. As the relation between the original bosons and the elementary excitations are connected by the paraunitary matrix $T_{\mathbf{k}}$, the interference of the edge mode and the condensate wavefunction is predicted to form a density wave for the original bosons along the edge~\cite{suppl}, by the same mechanism discussed in  Ref.~\cite{Furukawa2015}.

{\it Conclusion.} We study the elementary excitations of the staggered chiral $p_x\pm ip_y$ orbital superfluid realized in the Hamburg experiments~\cite{Wirth2011, Olschlager2013, Kock2015}. We find that for the experimentally implemented lattice configuration, the elementary excitations are not topologically gapped due to the presence of the combined symmetries ($\Theta_{\pm}\mathcal{T}$), whose topological classification falls
into the BDI class~\cite{Hasan2010} after taking into account the intrinsic particle-hole symmetry of $\tau_z\mathcal{H}(\mathbf{k})$. Four Dirac cones are found in the higher lying excitation spectrum. In constrast, when a bias square lattice potential imbalances the $p_x+ip_y$ and $p_x-ip_y$ components, a striking, unexpected topological phase transition appears. The resulting superfluid state supports topological gapped bulk elementary excitations accompanied by in-gap topologically protected edge excitations. This finding extends current research on topological phases of fermionic atoms and electrons to weakly interacting Bose atoms. Most significantly, the chiral orbital order and the topological elementary excitations we discuss are driven by the many-body interaction, which is in contrast to those requiring single-particle Bloch bands with topological character.

This work is supported by NSFC (No.~11574100) (Z.-F.X), 2013CB922004 of the National Key Basic Research Program of China, and by NSFC (No.~91121005, No.~91421305) (L.Y.), and U.S. ARO (W911NF-11-1-0230), AFOSR (FA9550-16-1-0006), the Charles E. Kaufman Foundation and The Pittsburgh Foundation, Overseas Collaboration Program of NSF of China (No. 11429402) sponsored by Peking University, and National Basic Research Program of China (No. 2012CB922101) (W.V.L.). A.H. acknowledges support by DFG-SFB925 and the Hamburg centre of ultrafast imaging (CUI). Z.-F.X. is also supported in part by the National Thousand-Young-Talents Program.

\clearpage
\onecolumngrid
\renewcommand\thefigure{S\arabic{figure}}
\setcounter{figure}{0}

{
\center \bf \large
Supplemental Material\\
\vspace*{0.25cm}
}

\newcommand{\mysetlabel}[2]{ \newcounter{maintextfig} \setcounter{maintextfig}{#2} \addtocounter{maintextfig}{-1}
\refstepcounter{maintextfig} \label{#1}}

\mysetlabel{fig3}{3}

In this supplementary material, we provide additional
details on (A) single-particle energy spectra, (2) methods for the derivation of Bogoliubov excitations, (C) two alternative schemes to open a bulk gap, (D) alignment of two optical lattice potentials, and (E) detection of edge states by density waves.

\vspace{0.1in}
\centerline{\bf (A) Single-particle energy spectra}
\vspace{0.1in}

\begin{figure}[H]
\centering
\includegraphics[width=0.6\linewidth]{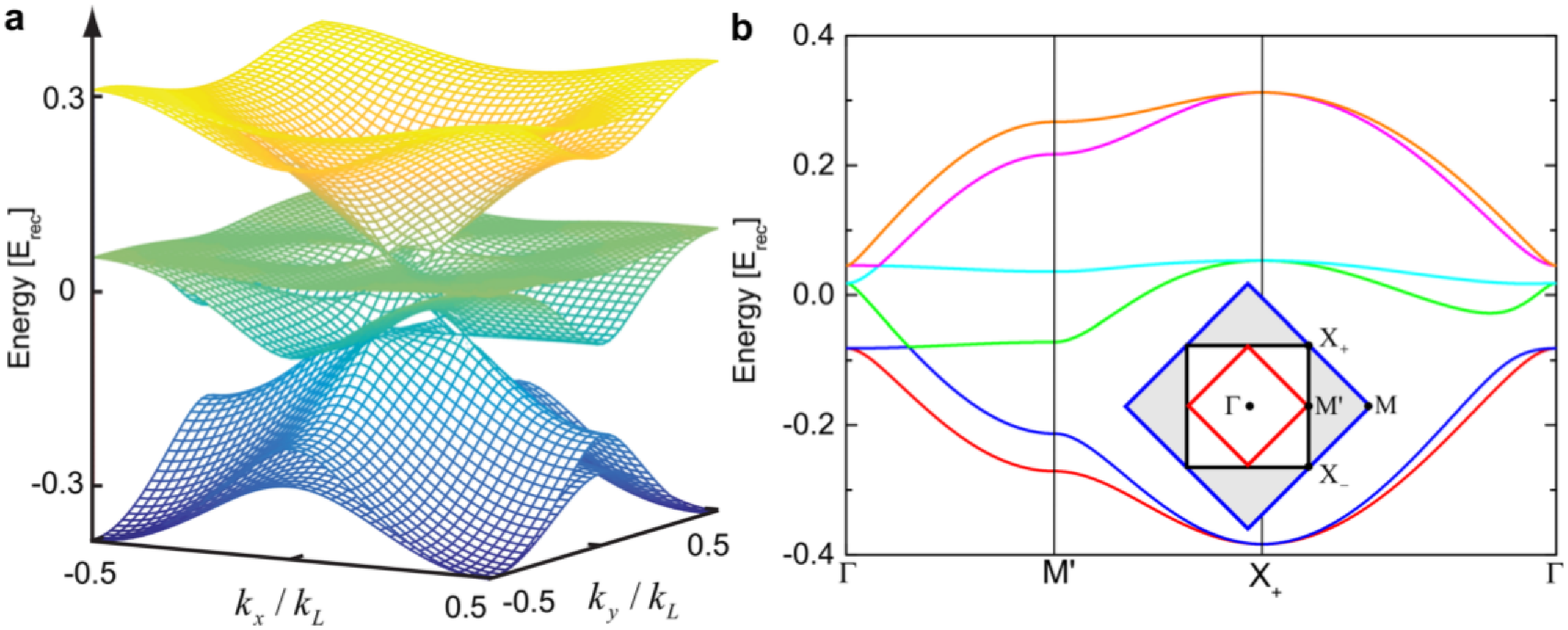}
\caption{Energy spectra of the single-particle Hamiltonian $H_0$ of the Methods section. (a) Single-particle energy spectra over the whole Brillouin zone. (b) Single-particle energy spectra along the high-symmetric lines. The parameters used here are the same as that used in the Fig.~2(d) of the main text where $V_S\ne 0$.}
\label{sfig1}
\end{figure}

The presence of the $V_S$ lattice potential induces a population imbalance between the $p_x+ip_y$ and $p_x-ip_y$ components of the chiral  $p_x\pm ip_y$ superfluid, which in the presence of interactions leads to a topological bulk gap of the Bogoliubov excitations close to the Dirac points. Here, we confirm that this topological bulk gap is indeed induced by the interaction. To clarify this point, we plot the single-particle energy spectra of the Hamiltonian $H_0$ in Fig.~\ref{sfig1}. It clearly shows that even though the bias potential $V_S$  generates an on-site energy difference for the $A1$ and $A2$ lattice sites, there is no bulk gap in the energy spectra, which validates our arguments.

\vspace{0.1in}
\centerline{\bf (B) Methods for the derivation of Bogoliubov excitations}
\vspace{0.1in}

\begin{figure}[H]
\centering
\includegraphics[width=0.6\linewidth]{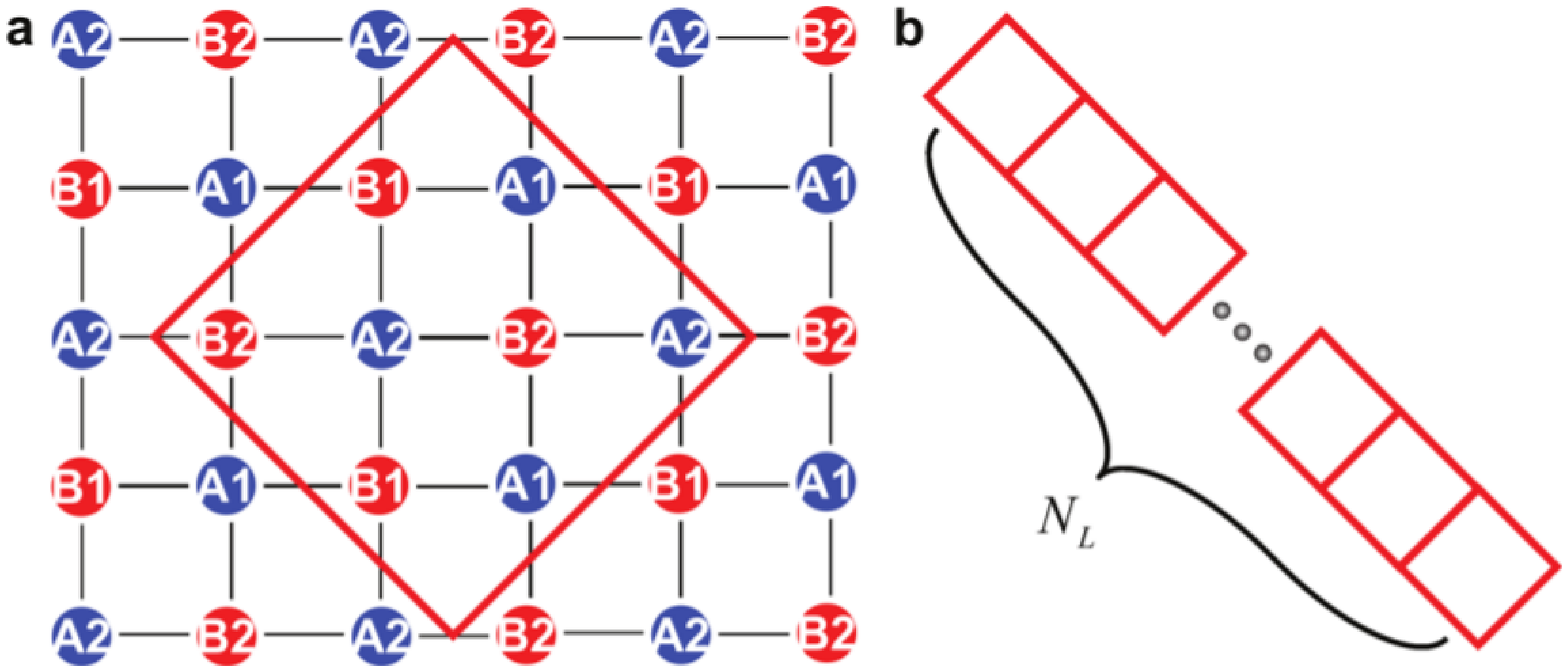}
\caption{Finite system. (a) Schematic diagram of the lattice structure. One unit cell of the order parameter is denoted by red solid square. (b) Finite system used in numerics to obtain the chiral edge states. It's periodic (or finite) along the $x+y$ (or $x-y$) direction. One unit cell of the finite system contains $N_L$ (usually 40 in our calculations) copies of red squares shown in (a).}
\label{sfig2}
\end{figure}

We arrange the 12 orbitals per unit cell of the order parameter in a specific order, and define their annihilation operators as $\hat{b}_{j_{u}\ell}$ ($\ell=1,...,12$ and $j_{u}$ characterizes different unit cells). In the reduced first Brillouin Zone, $[-\pi/2\sqrt{2}a,\pi/2\sqrt{2}a]\otimes[-\pi/2\sqrt{2}a,\pi/2\sqrt{2}a]$, bosons are now condensed at zero crystal momentum. We apply a Fourier transformation to the  operators defined by
\begin{eqnarray}
\hat{b}_{j_u\ell}=\sum\limits_{\mathbf{k}\in\rm RFBZ}\hat{b}_{\mathbf{k}\ell}e^{i\mathbf{k}\cdot\mathbf{r}_{j_{u}\ell}},
\hat{b}_{\mathbf{k}\ell}=\frac{1}{N_u}\sum\limits_{j_u}\hat{b}_{j_u\ell} e^{-i\mathbf{k}\cdot\mathbf{r}_{j_{u\ell}}},
\end{eqnarray}
where $\mathbf{r}_{j_u\ell}$ denote the position of the orbitals, $N_u$ is the number of the unit cells, and $[\hat{b}_{j_u\ell},\hat{b}^{\dagger}_{j'_u\ell'}]=\delta_{j_uj'_u}\delta_{\ell\ell'}$, $[\hat{b}_{\mathbf{k}\ell},\hat{b}^{\dagger}_{\mathbf{k}'\ell'}]=(1/N_u)\delta_{\mathbf{k}\mathbf{k}'}\delta_{\ell\ell'}$.
To obtain the ground state, we first diagonalize the single-particle Hamiltonian $\hat{H}_0=N_u\sum_{\mathbf{k}}\mathcal{B}_{\mathbf{k}}^{\dagger}\mathcal{H}_0(\mathbf{k}) \mathcal{B}_{\mathbf{k}}$ as $\mathcal{D}_{\mathbf{k}}^{\dagger}\mathcal{H}_0(\mathbf{k})\mathcal{D}_{\mathbf{k}}= \mathcal{E}_{\mathbf{k}}$, where $\mathcal{B}_{\mathbf{k}}\equiv(\hat{b}_{\mathbf{k}1},...,\hat{b}_{\mathbf{k}12})^T$.
With bosons condensed at $\mathbf{k}=\mathbf{0}$, the ground-state wave function is given by $|\psi\rangle=(1/\sqrt{N!})\big\{\sum_{\ell}\psi_{\mathbf{0}\ell} \sqrt{N_u}(\mathcal{D}_{\mathbf{0}}^{\dagger}\mathcal{B}_{\mathbf{0}})^{\dagger}_{\ell}\big\}^N|\rm vac\rangle$, where $N$ is the total atom number. Minimizing the energy functional $\langle\psi|\hat{H}|\psi\rangle$ under the condition of a fixed atom number by the method of simulated annealing, we obtain $\langle b_{\mathbf{0}\ell}\rangle= \langle \psi|\hat{b}_{\mathbf{0}\ell}|\psi\rangle=\sqrt{N/N_u} (\mathcal{D}_{\mathbf{0}}\psi_{\mathbf{0}})_{\ell}$, where $\psi_{\mathbf{0}}=(\psi_{\mathbf{0}1},...,\psi_{\mathbf{0}12})^T$.
The orbital-exchange interaction fixes the relative phase between the $p_x$ and $p_y$ orbitals to be $\pm\pi/2$, which leads to a nonzero local orbital angular momentum.

We then follow the number-conserving approach to obtain the Bogoliubov excitations by replacing the bosonic annihilation operators $\hat{b}_{\mathbf{0}\ell}$ with $\big(N/N_u-\sum_{\mathbf{k}\ell'} \hat{b}_{\mathbf{k}\ell'}^{\dagger}\hat{b}_{\mathbf{k}\ell'}\big)^{1/2} (\mathcal{D}_{\mathbf{0}}\psi_{\mathbf{0}})_{\ell}$ for the Hamiltonian. Keeping up to the quadratic term of the operators, the Hamiltonian is rewritten as $\hat{H}=(N_u/2)\sum_{\mathbf{k}\ne0}(\mathcal{B}_{\mathbf{k}}^{\dagger}, \mathcal{B}_{-\mathbf{k}}^T)\mathcal{H}({\mathbf{k})(\mathcal{B}_{\mathbf{k}}, \mathcal{B}_{-\mathbf{k}}^{\dagger T}})^T $, where
\begin{eqnarray}
\mathcal{H}(\mathbf{k})=
\left(
\begin{array}{cc}
  \mathcal{M}_{\mathbf{k}} & \mathcal{N}_{\mathbf{k}} \\
  \mathcal{N}_{-\mathbf{k}}^{*} & \mathcal{M}_{-\mathbf{k}}^{*}
\end{array}
\right),
\end{eqnarray}
with $\mathcal{M}_{\mathbf{k}}$ hermitian. To satisfy the bosonic commutation relation, the Bogoliugov-de Gennes Hamiltonian is diagonalized by a paraunitary matrix as
$T^{\dagger}_{\mathbf{k}}\mathcal{H}(\mathbf{k})T_{\mathbf{k}}=E_{\mathbf{k}}$,
where $T_{\mathbf{k}}^{\dagger}\tau_zT_{\mathbf{k}}=\tau_z$ ($[\tau_z]_{ij}=\delta_{ij}$ if $i=1,...,12$; $-\delta_{ij}$ otherwise) leading to $T^{-1}_{\mathbf{k}}=\tau_zT^{\dagger}_{\mathbf{k}}\tau_z$.
The corresponding eigenvalues are contained in the diagonal terms of the diagonal matrix $E_{\mathbf{k}}$.

For the finite system as shown in Fig.~\ref{sfig2}, we still assume that the order parameter of the condensate takes the same value as that for the infinite system. This assumption will lead to a tiny gap close to zero energy for the excitations. Numerically, we slightly change the chemical potential to smear out the tiny gap. The needed chemical change decreases as we enlarge the system. All in all, this artificial technique does not change the topological features of the highly excited Bogoliubov excitations.

\begin{figure}[tbp]
\centering
\includegraphics[width=0.6\linewidth]{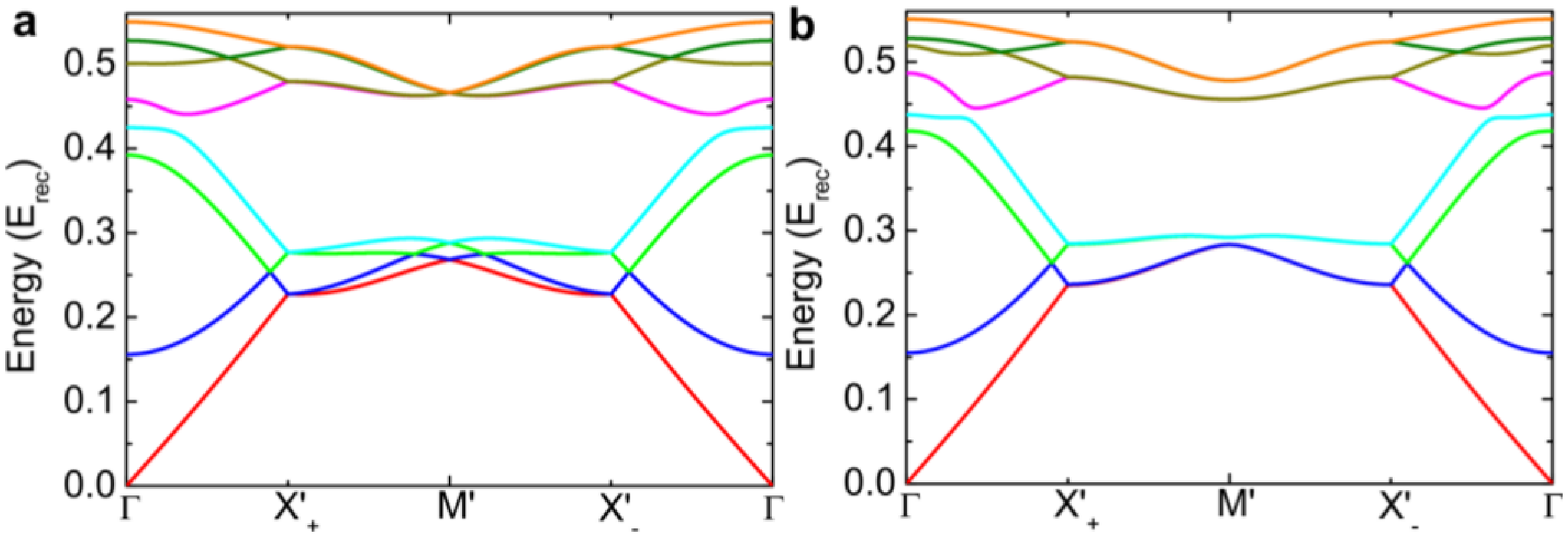}
\caption{Bogoliubov excitation spectra. (a) Bogoliubov excitation spectra for $V_S=0$ but with an extra coupling between $p$ orbitals in next-nearest neighbor sites along the high-symmetric lines. The parameters used in numerics are $J_{\rm sp}=0.12\, E_{\rm rec}$, $J_{\parallel}=0.07\, J_{\rm sp}$, $J_{\perp}=0.15\, J_{\rm sp}$, $\Delta V_{\rm sp}=0.3\, J_{\rm sp}$, $U_s=0.24\, E_{\rm rec}$, $U_p=0.12\, E_{\rm rec}$, $\Delta_{\rm pp}=0$, and $\Delta J_{\rm sp}=0.4\, J_{\rm sp}$. The mean boson density is one particle per orbital. (b) Bogoliubov excitation spectra for $V_S=0$ and an on-site rotation. The parameters used in numerics are the same as (a) except $\Delta J_{\rm sp}=0$ and $\Omega_z=0.1\,J_{\rm sp}$.}
\label{sfig3}
\end{figure}

\begin{figure}[H]
\centering
\includegraphics[width=0.8\linewidth]{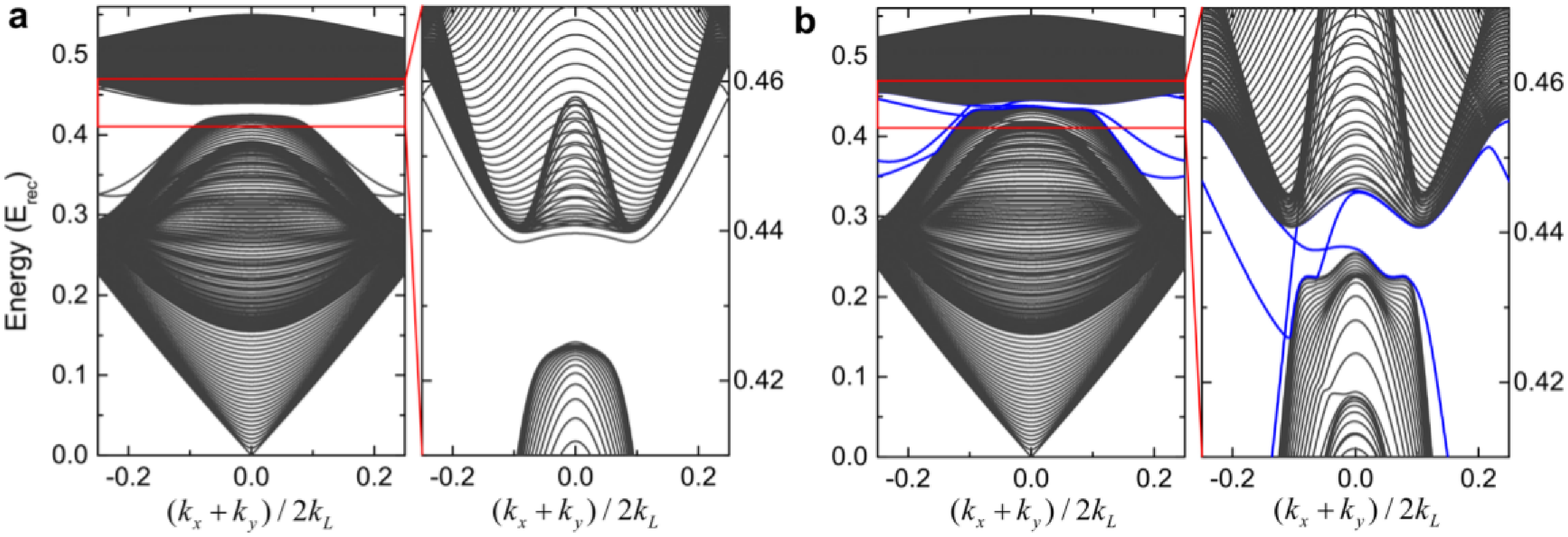}
\caption{Edge states. The corresponding energy spectra of the two different cases shown in Fig.~\ref{sfig3} for the finite system with a stripe geometry shown in Fig.~\ref{sfig1}(b).}
\label{sfig4}
\end{figure}

\vspace{0.1in}
\centerline{\bf (C) Two alternative schemes to open a bulk gap at Dirac points}
\vspace{0.1in}

\begin{figure}[tpb]
\centering
\includegraphics[width=0.6\linewidth]{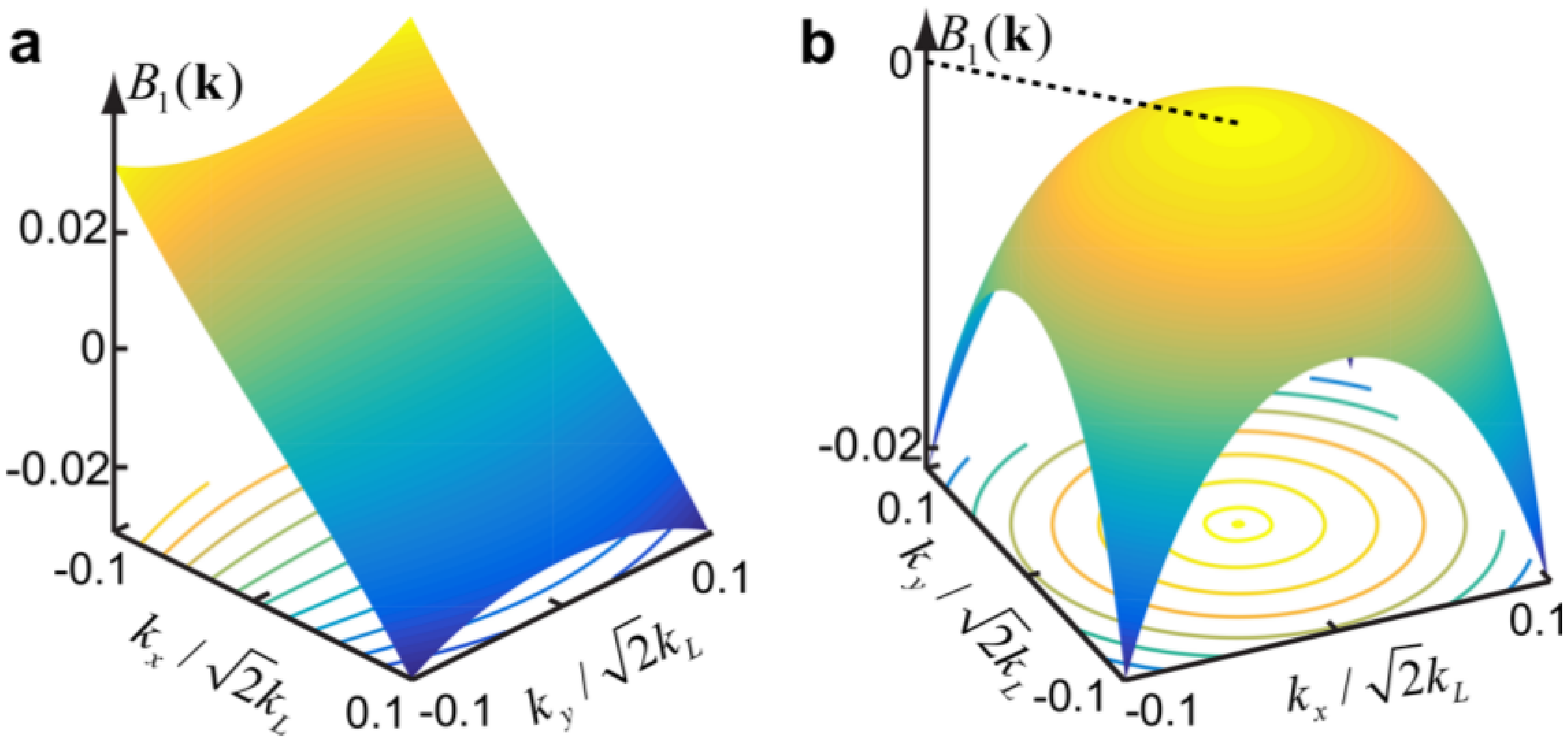}
\caption{Phonon mode. The corresponding Berry curvature for the phonon mode of two different cases shown in Fig.~\ref{sfig3}.}
\label{sfig5}
\end{figure}

In the main text, we discuss two alternative schemes to open a bulk gap close to Dirac points. The first one is realized by introducing an extra noncentrosymmetric optical lattice potential~\cite{sTarruell2012}, which induces a change to the nearest-neighbor hopping between $s$ and $p_y$ orbitals along $y$-axis. However, this method does not break the $\Theta_{\pm}\mathcal{T}$ symmetry. Thus, it only opens a topological trivial bulk gap, which is confirmed numerically by finding that $|\sum_{j=5}^8\mathcal{C}_j|=0$ and no topological protected edge modes arise for the finite system shown in Fig.~\ref{sfig4}(a). Nevertheless, we find that the Berry curvature for phonon modes is nonzero although it takes different sign depending on the value of momentum, as shown in Fig.~\ref{sfig5}(a). The second one is realized by introducing an on-site rotation~\cite{sGemelke2010,sZhang2011}, which breaks the time-reversal and $\Theta_{\pm}\mathcal{T}$ symmetries. In this case, a topological nontrivial gap opens up with $|\sum_{j=5}^8\mathcal{C}_j|=2$, and topological protected edge modes appear in the bulk gap as shown in Fig.~3(b). In the meantime, the Berry curvature for the phonon mode is nonzero and is of a fixed sign as shown in Fig.~\ref{sfig5}(b), which should also give rise to a phonon Hall effect as in the model system with $V_S$ discussed in the main text.

\vspace{0.1in}
\centerline{\bf (D) Alignment of two optical lattice potentials}
\vspace{0.1in}

\begin{figure}[H]
\centering
\includegraphics[width=0.6\linewidth]{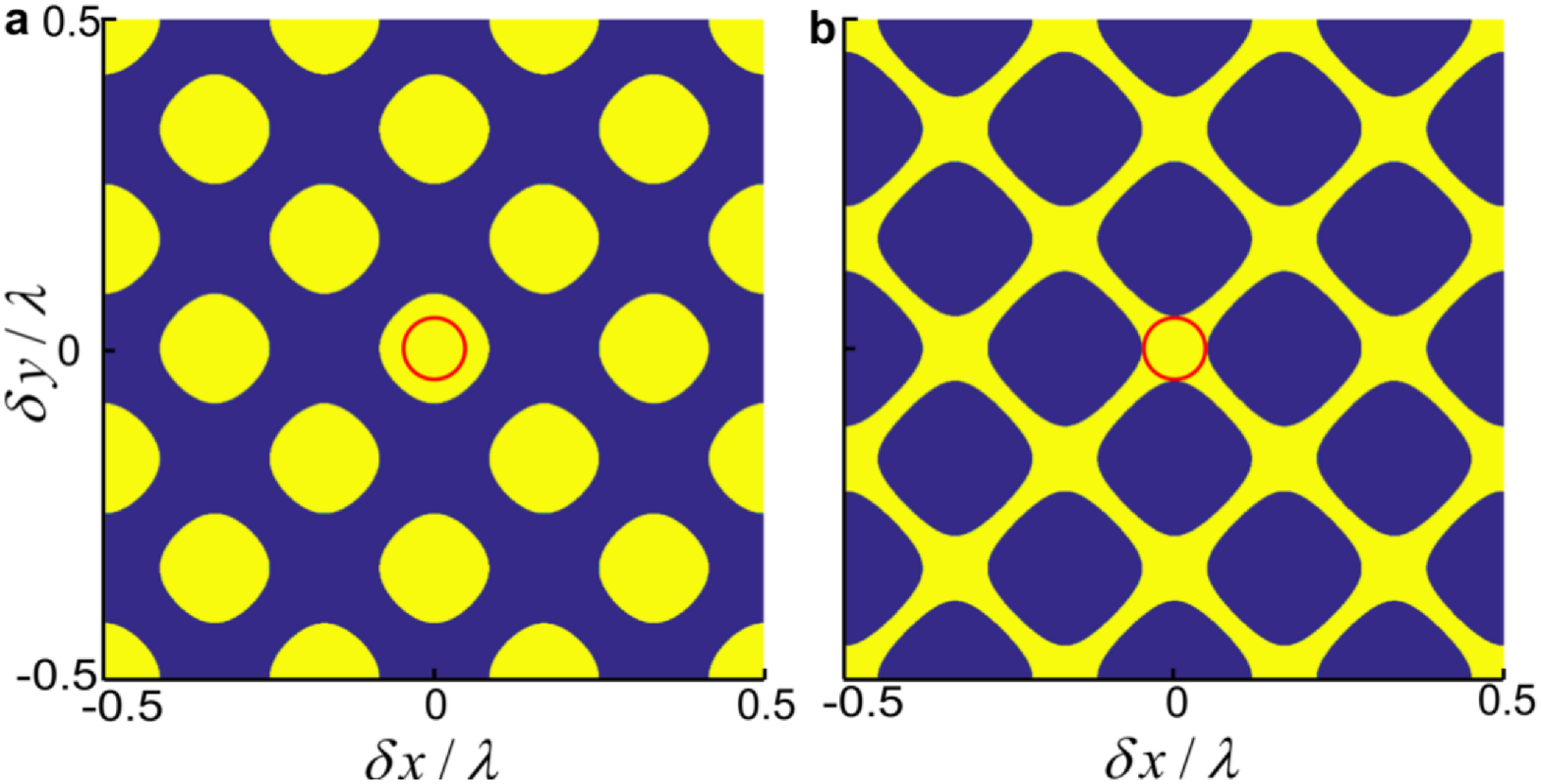}
\caption{Staggered on-site energy induced by a shifted $V_S$. In realistic experiments, the center of $V_S$ may be shifted away from the center of $V_H$, which is described by $V_S(x,y)=V_2\big[\cos^2\big(3k_L(x+\delta x)/2\big)+\cos^2\big(3k_L(y+\delta y)/2\big)\big]$. The mismatch of two lattice centers will change the induced on-site energy differences for p-orbitals at $A1$ and $A2$ sites, which is crucial for generating topological Bogoliubov excitations. (a) On-site energy change for the p-orbitals. The yellow (or blue) color denotes the region where $|V_S(A1)-V_S(A2)|\ge |V_2|$ \big(or $|V_S(A1)-V_S(A2)|<|V_2|$\big). (b) On-site energy change for the s-orbitals. The yellow (or blue) color denotes the region where $|V_S(B1)-V_S(B2)|\le 2|V_2|/5$ \big(or $|V_S(B1)-V_S(B2)|>2|V_2|/5$\big). Here, $V_S(A1,B1, A2, B2)$ indicates the value of extra potential $V_S$ at $A1$, $B1$, $A2$, and $B2$ sites, respectively. The diameter of the red circle is $100\, \rm nm$.}
\label{sfig6}
\end{figure}

Here, we consider a realistic experimental setup, where the center of the bias lattice potential $V_S$ deviates from the center of the main lattice potential $V_H$. The purpose of adding a bias lattice potential $V_S$ is to induce a population imbalance between $p$ orbitals at $A1$ and $A2$ sublattices. A weak potential can satisfy our requirement. Thus, the main effect of $V_S$ is to change the on-site energy for $s$ and $p$ orbitals. Figure~\ref{sfig6} shows the change of the on-site energy difference for $p$ orbitals (\textbf{a}) or $s$ orbitals (\textbf{b}) induced by a shifted $V_S$. We infer that if the center mismatch between two lattices is not large, e.g. $\sim 50\,$nm, the induced on-site energy difference between $p$ orbitals at $A1$ and $A2$ lattices is still comparable to the perfectly matched case. This requirement can be readily realized with standard technology, as pointed out in the main text. Lattices with two components spatially positioned relative to each other with high precision have been also used in a number of previous experiments \cite{sJo2012,sTaie2015}.

\begin{figure}[H]
\centering
\includegraphics[width=0.8\linewidth]{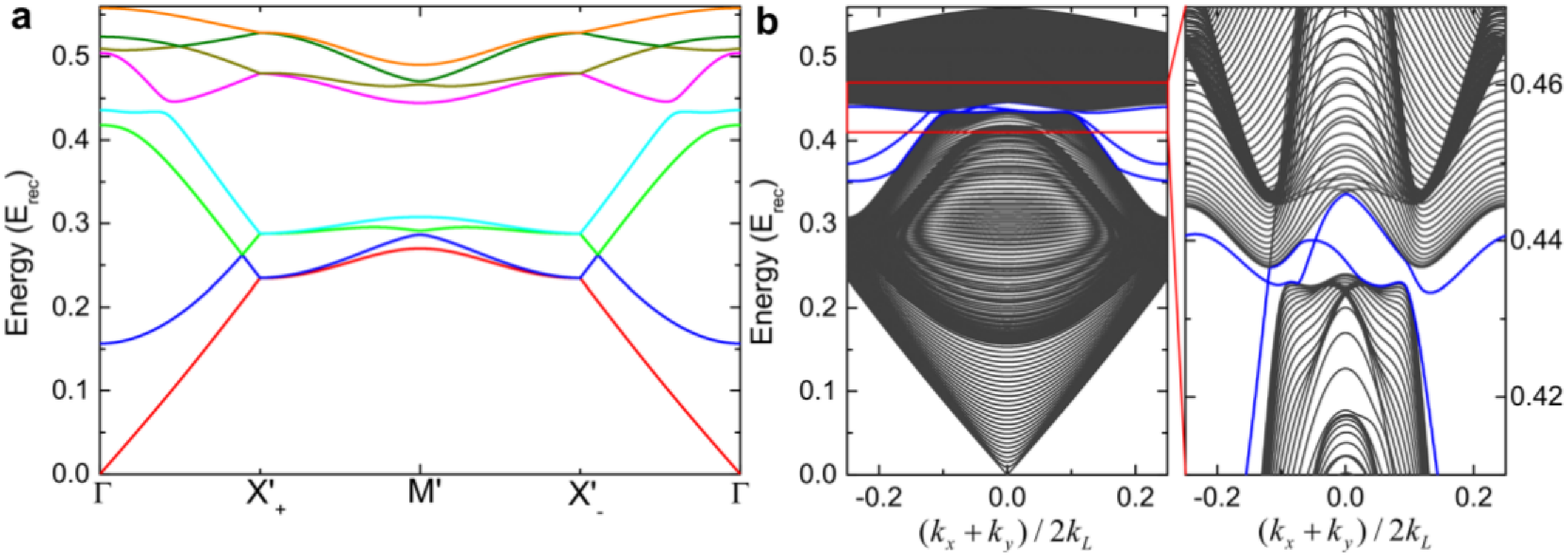}
\caption{Bogoliubov excitation spectra and edge modes. Here, we assume that $V_S$ induces on-site energy differences both for $s$ orbitals at $B1$ and $B2$ sites and $p$ orbitals at $A1$ and $A2$ sites. (a) Bogoliubov excitation spectra. (b) Edge modes for a finite system. Parameters are the same as that used in Fig.~3, except that the on-site energy difference is also introduced for the $s$ orbitals at $B1$ and $B2$ lattice sites.}
\label{sfig7}
\end{figure}

Since the on-site energy difference for $p$ orbitals at $A1$ and $A2$ sites is introduced by $V_S$, the population imbalance between $p_x+ip_y$ and $p_x-ip_y$ is generated. In our numerics, we confirm that such imbalance leads to topological Bogoliubov excitations. The on-site energy difference for $s$ orbitals at $B1$ and $B2$ sites is found irrelevant for the topological Bogoliubov excitations. We have performed calculations to verify that even if the on-site energy difference is introduced only for the $s$ orbitals, four Dirac points still persist. In contrast, no matter whether such difference is introduced for $B1$ and $B2$ sites or not, the presence of any on-site energy difference between $p$ orbitals at $A1$ and $A2$ sites is found to open a topologically nontrivial bulk gap close to Dirac points. Figure~\ref{sfig7} demonstrates the results for the case where there are on-site energy differences both for $s$ orbitals at $B1$ and $B2$ sites and for $p$ orbitals at $A1$ and $A2$ sites. In conclusion, the slight mismatch of two lattice centers will not change our conclusion that topological Bogoliubov excitations can be induced by turning on $V_S$.

\vspace{0.1in}
\centerline{\bf (E) Detection of edge states by density waves}
\vspace{0.1in}

To detect the edge states, we propose to coherently transfer a portion of the condensate into an edge mode by stimulated Raman transitions. We consider a finite system with an open (or periodic) boundary along $x+y$ (or $x-y$) direction as shown in Fig.~\ref{sfig1}. The sharp boundaries here can be realized by putting the system in a box trap~\cite{sGaunt2013}. Following the standard method, we obtain a Bogoliubov-de Gennes Hamiltonian $\mathcal{H}(\mathbf{k})$ under the basis $(\mathcal{B}_{\mathbf{k}},\mathcal{B}_{-\mathbf{k}}^{\dagger T})^T$ where $\mathcal{B}_{\mathbf{k}}=(\dots,\hat{b}_{\mathbf{k}\ell},\dots)^T$ and $\ell$ is used to label all considered $s$ and $p$ orbitals in one unit cell of the finite system as shown Fig.~\ref{sfig1}(b). Recall that there are a total of $12\times N_L$ elements in $\mathcal{B}_{\mathbf{k}}$ with $N_L=40$ for the finite system we consider. Thus $\ell=1,\dots, 12 N_L$. Applying a transformation $(\mathcal{B}_{\mathbf{k}},\mathcal{B}_{-\mathbf{k}}^{\dagger T})^T=T_{\mathbf{k}}(\mathcal{A}_{\mathbf{k}},\mathcal{A}_{-\mathbf{k}}^{\dagger T})^T$ where $\mathcal{A}_{\mathbf{k}}\equiv(\dots,\hat{a}_{\mathbf{k}\ell},\dots)^T$, the Hamiltonian is then diagonalized as $T_{\mathbf{k}}^{\dagger}\mathcal{H}(\mathbf{k})T_{\mathbf{k}}=E_{\mathbf{k}}$.
The $\ell$-th column of $T_{\mathbf{k}}$ is labeled as  $(u_{\mathbf{k}\ell}^T,v_{\mathbf{k}\ell}^T)^T$ where both $u_{\mathbf{k}\ell}$ and $v_{\mathbf{k}\ell}$ are $12N_L\times 1$ matrices.

\begin{figure}[H]
\centering
\includegraphics[width=0.6\linewidth]{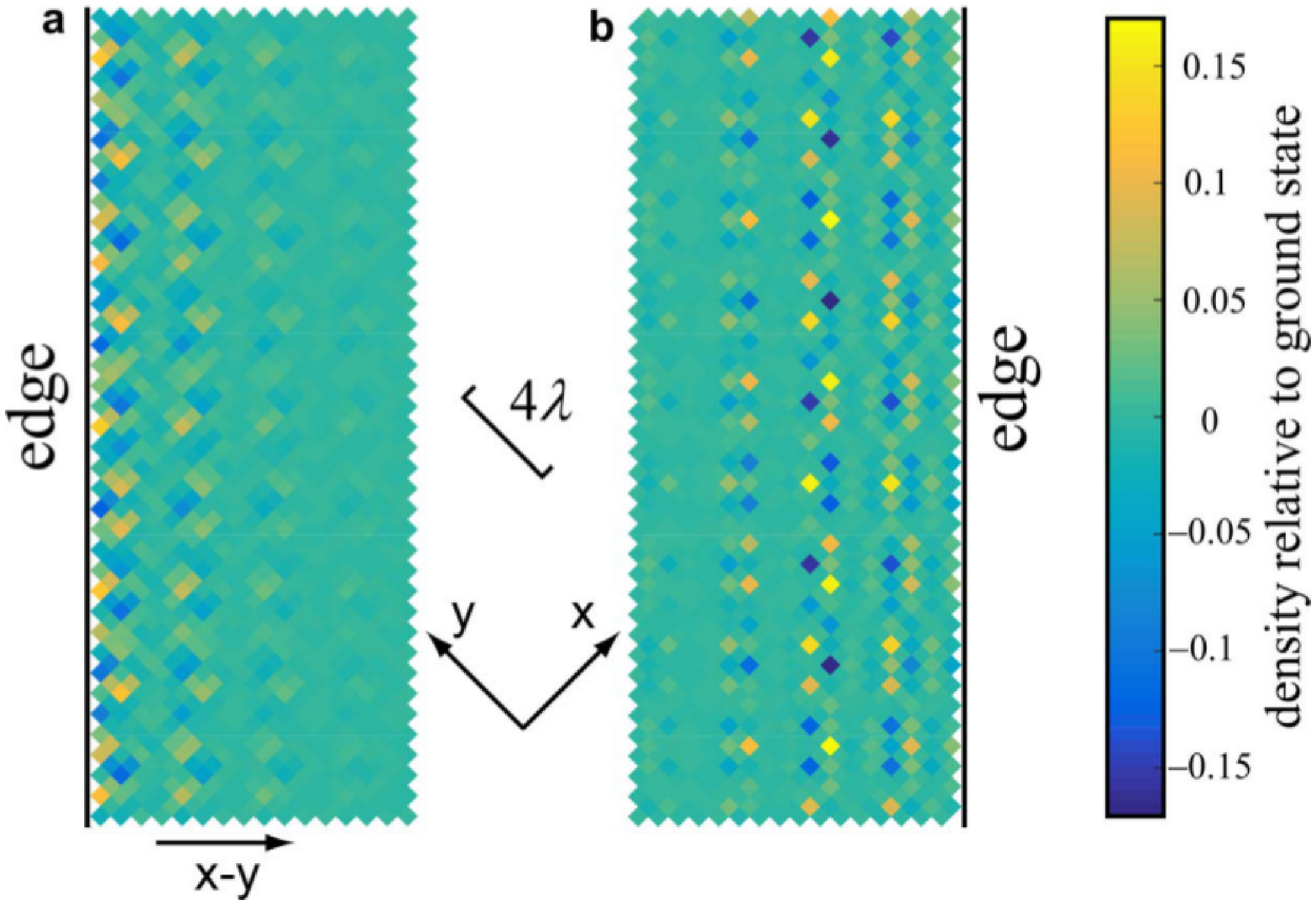}
\caption{\textbf{Interference pattern of the edge mode and the ground-state condensate wavefunction.} A small portion (about 5 percent) of the condensate is suddenly transferred into the edge state with $k_x=k_y=0.115\,k_L$ and quasi energy $0.4375\, E_{\rm rec}$ in \textbf{a} and $0.4402\,E_{\rm rec}$ in \textbf{b}. The color shows the density relative to that of the ground state $|\langle \hat{b}_{\mathbf{0}\ell}\rangle+\langle \hat{b}_{\mathbf{q}\ell}\rangle e^{i\mathbf{q}\cdot\mathbf{r}}+\langle \hat{b}_{-\mathbf{q}\ell}\rangle e^{-i\mathbf{q}\cdot\mathbf{r}}|^2-|\langle \hat{b}_{\mathbf{0}\ell}\rangle|^2$. The center of each square pixel corresponds to a lattice site. Here, only the region close to the left or the right edge is shown. The equivalent of charge density waves is observed in the figures near the respective edge, indicating the existence of topologically protected edge modes absent in the bulk.}
\label{sfig8}
\end{figure}

Experimentally, after transferring a small portion of the condensate into the $\ell'$-th Bogoliubov excitation band at momentum $\mathbf{q}$, we have $\langle a_{\mathbf{q}\ell'}\rangle\ne 0$. Straight forward calculations show $\langle\hat{b}_{\mathbf{q}\ell}\rangle=(u_{\mathbf{q}\ell})_{\ell'} \langle\hat{a}_{\mathbf{q}\ell'}\rangle$ and $\langle\hat{b}^{\dagger}_{-\mathbf{q}\ell}\rangle=(v_{\mathbf{q}\ell})_{\ell'} \langle\hat{a}_{\mathbf{q}\ell'}\rangle$, where $(u_{\mathbf{q}\ell})_{\ell'}$ and $(v_{\mathbf{q}\ell})_{\ell'}$ denote the $\ell'$-th element of $u_{\mathbf{q}\ell}$ and $v_{\mathbf{q}\ell}$, respectively. Including the ground-state condensate where bosons are condensed at $k_x=k_y=0$ with $\langle b_{\mathbf{0}\ell}\rangle\ne0$, the condensate wavefunction $\psi(\mathbf{r})$ immediately after transferring a small portion into an excitation mode is given by $\langle \hat{b}_{\mathbf{0}\ell}\rangle+\langle \hat{b}_{\mathbf{q}\ell}\rangle e^{i\mathbf{q}\cdot\mathbf{r}}+\langle \hat{b}_{-\mathbf{q}\ell}\rangle e^{-i\mathbf{q}\cdot\mathbf{r}}$. The interference of the edge state and the background condensate provides a clear signal for the former (Fig.~\ref{sfig8}). While atoms here are charge neutral particles, the interference clearly shows the equivalent of charge density waves often found in electronic solid-state materials, which provides the evidence of the topologically protected edge modes existing in the bulk gap.

\begin{thebibliography}{999}

\bibitem{Hasan2010}
M. Z. Hasan and C. L. Kane, Rev. Mod. Phys. {\bf 82}, 3045 (2010).

\bibitem{Qi2011}
X. L. Qi and S.-C. Zhang, Rev. Mod. Phys. {\bf 83}, 1057 (2011).

\bibitem{Aidelsburger2013}
M. Aidelsburger, M. Atala, M. Lohse, J. T. Barreiro, B. Paredes, and I.  Bloch, Phys. Rev. Lett. {\bf 111}, 185301 (2013).

\bibitem{Miyake2013}
H. Miyake, G. A. Siviloglou, C. J. Kennedy, W. C. Burton and W. Ketterle, Phys. Rev. Lett. {\bf 111}, 185302 (2013).

\bibitem{Aidelsburger2015}
M. Aidelsburger, M. Lohse, C. Schweizer, M. Atala, J. T. Barreiro, S. Nascimb\`ene, N. R. Cooper, I. Bloch, and N. Goldman, Nature Physics {\bf 11}, 162-166 (2015).

\bibitem{Mancini2015}
M. Mancini, G. Pagano, G. Cappellini, L. Livi, M. Rider, J. Catani, C. Sias, P. Zoller, M. Inguscio, M. Dalmonte, and L. Fallani, Science {\bf 349}, 1510 (2015).

\bibitem{Stuhl2015}
B. K. Stuhl, H.-I. Lu, L. M. Aycock, D. Genkina, and I. B. Spielman, Science {\bf 349}, 1514 (2015).

\bibitem{Jotzu2014}
G. Jotzu, M. Messer, R. Desbuquois, M. Lebrat, T. Uehlinger, D. Greif, and T. Esslinger, Nature {\bf 515}, 237 (2014).

\bibitem{Lin2011}
Y.-J. Lin, K. Jim{\'e}nez-Garcia, and I. B. Spielman, Nature {\bf 471}, 83 (2011).

\bibitem{Struck2011}
J. Struck, C. \"Olschl\"ager, R. Le Targat, P. Soltan-Panahi, A. Eckardt, M. Lewenstein, P. Windpassinger, and K. Sengstock, Science {\bf 333}, 996 (2011).

\bibitem{Parker2013}
C. V. Parker, L. C. Ha, and C. Chin, Nature Physics {\bf 9}, 769 (2013).

\bibitem{Volovik2003}
G. E. Volovik, The universe in a helium droplet, Oxford University Press, (2003).

\bibitem{Kallin2012}
C. Kallin, Rep. Prog. Phys. {\bf 75}, 042501 (2012).

\bibitem{Nayak2008}
C. Nayak, S. H. Simon, A. Stern, M. Freedman, and S. Das Sarma, Rev. Mod. Phys. {\bf 80}, 1083 (2008).

\bibitem{Engelhardt2015}
G. Engelhardt and T. Brandes, Phys. Rev. A {\bf 91}, 053621 (2015).

\bibitem{Furukawa2015}
S. Furukawa and M. Ueda, New J. Phys. {\bf 17}, 115014 (2015).

\bibitem{Bardyn2015}
C.-E. Bardyn, T. Karzig, G. Refael, and T. C. H. Liew, Phys. Rev. B {\bf 93}, 020502(R) (2016).

\bibitem{Wirth2011}
G. Wirth, M. \"Olschl\"ager, and A. Hemmerich, Nature Physics {\bf 7}, 147 (2011).

\bibitem{Olschlager2013}
M. \"Olschl\"ager, T. Kock, G. Wirth, A. Ewerbeck, C. M. Smith, and A. Hemmerich, New J. Phys. {\bf 15}, 083041 (2013).

\bibitem{Kock2015}
T. Kock, M. \"Olschl\"ager, A. Ewerbeck, W.-M. Huang, L. Mathey, and A. Hemmerich, Phys. Rev. Lett. {\bf 114}, 115301 (2015).

\bibitem{Maeno1994}
Y. Maeno, H. Hashimoto, K. Yoshida, S. Nishizaki, T. Fujita, J. G. Bednorz, and F. Lichtenberg, Nature {\bf 372}, 532 (1994).

\bibitem{Mackenzie2003}
A. P. Mackenzie and Y. Maeno, Rev. Mod. Phys. {\bf 75}, 657 (2003).

\bibitem{Hemmerich1991}
A. Hemmerich, D. Schropp, and T. W. H\"ansch, Phys. Rev. A {\bf 44}, 1910 (1991).

\bibitem{Isacsson2005}
A. Isacsson and S. M. Girvin, Phys. Rev. A {\bf 72}, 053604 (2005).

\bibitem{Liu2006}
W. V. Liu and C. Wu, Phys. Rev. A {\bf 74}, 013607 (2006).

\bibitem{suppl}
See Supplemental Material which includes
Refs.~\cite{sTarruell2012,sGemelke2010,sZhang2011,sJo2012,sTaie2015,sGaunt2013} for additional details on single-particle energy spectra, methods for the derivation of elementary excitations, two alternative schemes to open a bulk gap, alightment of two optical lattices, and edge-state detection.

\bibitem{Avron1983}
J. E. Avron, R. Seiler, and B. Simon, Phys. Rev. Lett. {\bf 51}, 51 (1983).

\bibitem{Shindou2013}
R. Shindou, R. Matsumoto, S. Murakami, and J.-i. Ohe, Phys. Rev. B {\bf 87}, 174427 (2013).

\bibitem{Neto2009}
A. H. Castro Neto, F. Fuinea, N. M. R. Peres, K. S. Novoselov, and A. K. Geim, Rev. Mod. Phys. {\bf 81}, 109 (2009).

\bibitem{Sun2012}
K. Sun, W. V. Liu, A. Hemmerich, S. Das Sarma, Nature Physics {\bf 8}, 67 (2012).

\bibitem{Atala2013}
M. Atala, M. Aidelsburger, J. T. Barreiro, D. Abanin, T. Kitagawa, E. Demler, and I. Bloch, Nature Physics {\bf 9}, 795 (2013).

\bibitem{Duca2015}
L. Duca, T. Li, M. Reitter, I. Bloch, M. Schleier-Smith, and U. Schneider, Science {\bf 347}, 288 (2015).

\bibitem{Xiao2010}
D. Xiao, M.-C. Chang, and Q. Niu, Rev. Mod. Phys. {\bf 82}, 1959-2007 (2010).

\bibitem{Li2014}
X. Li, A. Paramekanti, A. Hemmerich, and W. V. Liu, Nature Communications {\bf 5}, 3205 (2014).

\bibitem{Ernst2010}
P. T. Ernst, S. G\"otze, J. S. Krauser, K. Pyka, D.-S. L\"uhmann, D. Pfannkuche, and K. Sengstock, Nature Physics {\bf 6}, 56 (2010).

\bibitem{sTarruell2012}

L. Tarruell, D. Greif, T. Uehlinger, G. Jotzu, and T. Esslinger, Nature {\bf 483}, 302 (2012).

\bibitem{sGemelke2010}
N. Gemelke, E. Sarajlic, and S. Chu, arXiv:1007.2677.

\bibitem{sZhang2011}
M. Zhang, H.-h. Hung, C. Zhang, and C. Wu, Phys. Rev. A {\bf 83}, 023615 (2011).

\bibitem{sJo2012}
G.-B., Jo, J. Guzman, C. K. Thomas, P. Hosur, A. Vishwanath, and D. M. Stamper-Kurn, Phys. Rev. Lett. {\bf 108}, 045305 (2012).

\bibitem{sTaie2015}
S. Taie, H. Ozawa, T. Ichinose, T. Nishio, S. Nakajima, and Y. Takahashi, Sci. Adv. {\bf 1}, e1500854 (2015).

\bibitem{sGaunt2013}
A. L. Gaunt, T. F. Schmidutz, I. Gotlibovych, R. P. Smith, and Z. Hadzibabic, Phys. Rev. Lett. {\bf 110}, 200406 (2013).

\end{thebibliography}
\end{document}